\documentclass[twocolumn,preprintnumbers,amsmath,amssymb]{revtex4}
\usepackage{bm,graphicx,graphics,amsmath,amssymb,bm,epsfig,color,pifont}
\usepackage{euscript,tabularx,textcomp}
\usepackage[english]{babel}

\begin{document}
\title{Micromagnetic Simulations of Ferromagnetic Rings}
\author{Gabriel D. Chaves-O'Flynn} 
\affiliation{Department of Physics, New York University, 4 Washington Place, New York, New York 10003, USA}
\author{Ke Xiao} 
\affiliation{Department of Physics, New York University, 4 Washington Place, New York, New York 10003, USA}
\author{D.L. Stein} 
\affiliation{Department of Physics, New York University, 4 Washington Place, New York, New York 10003, USA}
\author{A.D. Kent}
\affiliation{Department of Physics, New York University, 4 Washington Place, New York, New York 10003, USA}

\date{09/12/07} 

\begin{abstract}
Thin nanomagnetic rings have generated interest for fundamental studies of
magnetization reversal and also for their potential in various
applications, particularly as magnetic memories. They are a rare example of
a geometry in which an analytical solution for the rate of thermally
induced magnetic reversal has been determined, in an approximation whose
errors can be estimated and bounded. In this work, numerical simulations of
soft ferromagnetic rings are used to explore aspects of the analytical
solution. The evolution of the energy near the transition states confirms
that, consistent with analytical predictions, thermally induced
magnetization reversal can have one of two intermediate states: either
constant or soliton-like saddle configurations, depending on ring size and
externally applied magnetic field. The results confirm analytical
predictions of a transition in thermally activated reversal behavior as
magnetic field is varied at constant ring size. Simulations also show that
the analytic one dimensional model continues to hold even for wide rings.
\end{abstract}

\maketitle

\section{Introduction}
The study of magnetization reversal in meso- and nanoscale ferromagnets is
motivated by its importance in information storage
applications. Ring-shaped thin film nanomagnets are particularly
interesting because the absence of sharp edges inhibits nucleation which
may precipitate undesired reversals. In addition, the geometry allows an
analytic solution - rare in micromagnetics - for thermally induced
reversal~\cite{martens1}.

The lowest energy configurations of a soft ferromagnetic ring are a
clockwise and counterclockwise circulation of the magnetization. The field
generated by a current $I$ flowing through the axis of the ring ($\hat
z$-direction) will lift the degeneracy between the two orientations.  For
an effectively one-dimensional ring (ring width small compared to the
average radius, so that external magnetic field strength within the ring is
approximately constant), the reversal from the metastable state to the
stable state can occur through either of two processes: a global rotation
of the magnetization (which we will call the constant saddle state), or
reversal initiated within a localized region and subsequent expansion
(instanton saddle)~\cite{martens1}. Which of these processes occurs depends
on both ring size and applied magnetic field. Using micromagnetic
simulations in OOMMF~\cite{donahue1999} we have confirmed this general
picture, even for more general two dimensional rings.

\begin{figure} 
 \centering 
\includegraphics[width=0.4\textwidth]{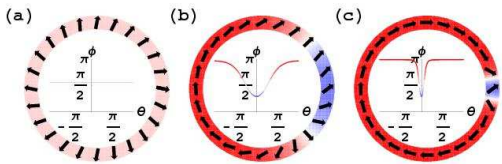}  
 \caption{Magnetization configurations for $h=0.2$. a) Constant saddle, b) Instanton Saddle at $l=12$, c) Instanton saddle at $l=60$.
 } 
 \label{fig.1} 
\end{figure}

The model introduced in~\cite{martens1} predicted that the magnetization will be
constrained to the film plane ($m_z=0$) and the transition states will
depend on both ring size and magnetic field. Important field and length
scales in the problem are: 
\begin{eqnarray}
\label{eq:scaled1}
h=\frac{H_e}{H_c}=\frac{H_e}{\frac{\mu_0 M_s}{\pi}\left(\frac{t}{
\Delta R} \right)\left | \hbox{ln}\left(\frac{t}{R}\right)\right |}\\
l=\frac{R}{\lambda}\sqrt{2\pi \left(\frac{t}{ \Delta R}\right)
\left|\hbox{ln}\left(\frac{t}{R}\right)\right|}\, .
\end{eqnarray} 
Here $M_s$ is the saturation magnetization, $t$ is the ring thickness,
$\Delta R$ is the ring width, $R$ is the average radius, $\lambda$ is the
exchange length, $H_e$ is the external magnetic field, and $H_c$ is the
field at which the metastable configuration becomes unstable.  For $l\le
2\pi\sqrt{1-h^2}$ the theory predicts a constant saddle, as shown in
Fig.~1a, whereas for $l>2\pi\sqrt{1-h^2}$, it predicts an instanton saddle
(Figs.~\ref{fig.1}b, c). Both of these saddle configurations are described
by a function $\phi_{h,l}(\theta)$~\cite{martens1}.
\section{Micromagnetic Simulations}

We simulated the magnetization dynamics of rings of mean radius $R=200$~nm
and thickness $t=2$~nm, for materials of exchange length $\lambda$ ranging
from $4$ to $40$~nm and constant saturation magnetization $M_s=8\times
10^5$~A/m (as in permalloy). The applied field has radial dependence $\vec
H(r)=\frac{I}{2\pi r}\hat \theta=\frac{R H_e}{r}\hat\theta$, where $H_e$ is
the magnitude of the magnetic field generated by the current $I$ at the
mean radius of the annulus.  We studied three different regimes for $l$:
(1) $l<2\pi$, where the transition state is predicted in~\cite{martens1} to
be the constant saddle configuration (Fig.~1a); (2) $l$ greater than but
close to $2\pi$ and (3) $l>>2 \pi$. The latter regime is relevant to recent
experiments on nanomagnetic
annuli~\cite{zhu2000,rothman2001,castano2003,castano2006,moneck2006,yang2007}.

The magnitude of the exchange stiffness (and therefore exchange length) was
varied so that $l$ lay in all three regimes: respectively, $l=6$, 12, and
60. Two different ring widths were studied: $\Delta R=40$ nm ($H_c=73.9$ mT)
and $\Delta R=100$ nm ($H_c=29.5$ mT). For different values of the parameter
$h$ the initial configurations (1a), (1b) and (1c) of the magnetization are
allowed to evolve following the Landau-Lifshitz-Gilbert equations:
\begin{equation}
\label{eq:LLG}
\frac{d\vec M}{dt}=-|\gamma| \vec M \times \vec H_{\sf eff} - \frac{|\gamma| \alpha}{M_s}\vec M\times(\vec M\times H_{\sf eff})
\end{equation}
where $\gamma$ is the gyromagnetic ratio and $H_{\sf eff}$ is the effective
magnetic field (i.e., including both the external field $H_e$ and
internal fields generated by the ring magnetization). The damping
coefficient $\alpha$ can be set to one without loss of generality.

We first conducted tests to determine how well the analytical solution
presented in~\cite{martens1} approximated the numerically determined
two-dimensional saddle state in a narrow ring.  We did this by initializing
the spin configuration at time $t=0$ using the instanton configuration
given by Eqs.~(8-11) of~\cite{martens1} with a fixed parameter $h$.  We
then allowed the system to evolve in time according to~(\ref{eq:LLG}) with
{\it different\/} values of the external field (of course, the external
field is fixed during any particular run).  A saddle state for a particular
external field $h_t$ must have the property that, for dynamics determined
by slightly higher fields, the system falls to the stable state; for
slightly lower fields, to the metastable state (or vice-versa).  Therefore,
for a fixed initial spin configuration defined by $h$ we varied the field
determining the dynamical evolution until a ``transition field'' $h_t$ was
found: for external fields less than $h_t$ the system would evolve to the
metastable configuration while for those greater than $h_t$ it would evolve
to the stable configuration.  In this way the saddle configuration can be
numerically determined.  Values of $h_t$ were obtained with a numerical
uncertainty of $\delta h_t=6 \times 10^{-3}$.

\section{Results and Discussion}
\label{sec:results}

The above discussion implies that the analytical solution is a good
approximation to the actual configuration if the field $h$ that determines
the initial instanton configuration is close to the numerically dtermined
$h_t$.  The results are shown in Fig.~2, which shows the the time evolution
of the energy. Typically, after a short transient the system arrives at a
configuration for which the energy stays almost constant in time before
decaying to either of the stable states. As $h$ gets closer to $h_t$, the
system maintains a constant energy for longer times. We interpret this
behavior as evidence that the analytically determined instanton predicted
in~\cite{martens1} is indeed a close approximation to the actual saddle
configuration. It can also be observed from Fig.~2 that at $h_t=0.21$ the
energy of the constant saddle is higher than that of the instanton saddle,
as predicted in~\cite{martens1}. The slope of the curve during the
subsequent relaxation depends on how fast the reversal propagates along the
ring: as $\lambda$ decreases the reversal time increases because both the
domain wall width decreases and the effective circumference $l$
increases. This reduces the magnitude of the slope of the curve $E(t)$
during the reversal.
\begin{figure} 
 \centering 
 \includegraphics[width=0.45\textwidth]{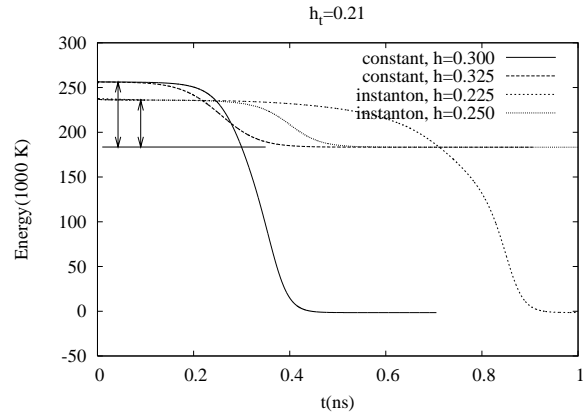} 
 \caption{Time evolution of the total energy in a field $h_t=0.21$, for the
 instanton ($h=\{0.225,0.25\}$) and constant saddle configurations
 ($h=\{0.3,0.325\}$) when $l$=12 and $\Delta R$=40 nm. These transition
 states are chosen to bracket the saddle state at the field of
 the simulation - the numerically determined $h_t$ is not the same as the
 parameter $h$ used to generate the saddle configurations from Eqs.~(8-11)
 of~\cite{martens1} (see text for a fuller explanation). The arrows
 represent the magnitudes of the activation energy $\Delta E(h_t=0.21)$ for
 each of the saddle states.}
 \label{fig. 2} 
\end{figure}

\begin{figure}
 \centering 
 \includegraphics[width=0.45\textwidth]{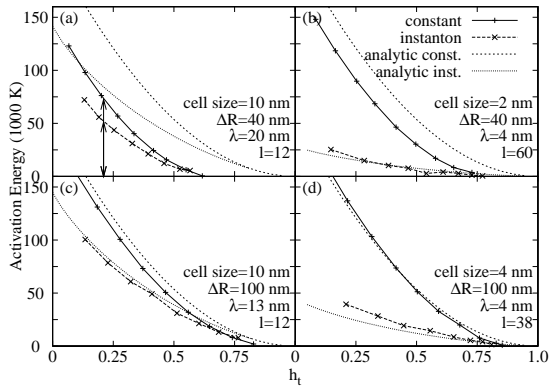} 
 \caption{Energy barrier dependence on 
$h_t$ for both saddle configurations
 in different regimes. Activation energies for $l=12, \Delta R=40$ nm at
$h_t=0.21$ are shown for comparison with Fig.~1. Data points represent
 $h=\{0.1,\ldots,0.9\}$. For comparison, theoretical lines are plotted for
 the activation energy. The crossing of the analytical curves in (b) and
 (d) is not visible due to the graphical resolution.}
 \label{fig. 3} 
\end{figure}

The reversal activation energy $\Delta E$ is summarized in
Fig.~\ref{fig. 3} as a function of $\Delta R$ and $\lambda$. Every datum
point corresponds to a step of $\Delta h=0.1$ in the parameter of each of
the saddle configurations. For fixed $l$, the energies of the instanton and
saddle configurations cross at $h_c=\sqrt{1-\left(\frac{2\pi}{l}
\right)^2}$; at higher fields the constant saddle is the lowest energy
transition state.  As predicted in~\cite{martens1}, the instanton saddle
has lower energy than the constant saddle for $h<h_c$, while for larger
fields the constant saddle has a lower energy (Fig.~3a, $h=0.57$). One
should observe, however, that the predicted external switching field is
larger than the numerically one. This is probably due to round-off from the
simulations, but more importantly from the fact that the theoretical result
is for a one-dimensional geometry whereas our simulations run in a (more
realistic) two-dimensional mesh. So it's surprising that the agreement
between the theoretical and numerically observed values for the switching
field {\it improves\/} when the width $\Delta R$ is increased from 40 nm to
100 nm; cf.~Figs.~3a-3c and 3b-3d. We believe this to be a consequence of
the increased number of active cells in the simulation for the wider ring
($\Delta R = $ 100 nm); it implies that the $1d$ solution is a good
approximation even for fairly wide two-dimensional ring geometries.

The theory presented in~\cite{martens1} predicts that the difference
between the activation energy of the constant saddle and the instanton
saddle increases as $\lambda$ decreases.  This is also confirmed by our
numerical simulations (compare Fig.~3a to 3b and 3c to 3d). This decrease
is due to fact that the activation energy for the constant saddle is
independent of $\lambda$ while the activation energy for the instanton
(whose size is order $\lambda$) is roughly proportional to $\lambda$. This
latter dependence of activation energy on $\lambda$ is in turn understood
by observing that the total energy of the ring is the result of three
contributions: Zeeman, exchange and magnetostatic energy. The constant
saddle, metastable and stable states all have the same exchange energy, so
the exchange energy does not contribute to the energy barrier of the
constant saddle. Both remaining terms (magnetostatic and Zeeman energy) are
independent of $\lambda$, so the constant saddle activation energy does not
change with $\lambda$. On the other hand, in most of the ring, the
metastable and instanton configurations are almost identical, so that their
energy difference is the result of a fluctuation in a small region of the
ring, whose size is of order $\lambda$. The barrier (both theoretical and
experimental) for the instanton saddle is therefore smaller in Figs.~3b and
3d than in Figs.~3a and 3c.

\section{Conclusion}

In these studies we have conducted numerical micromagnetic simulations of
the transition states for magnetization reversal in thin annular
nanomagnets to test the theory presented in~\cite{martens1}.
Numerical calculations are consistent with the prediction that in the
parameter space defined by $(h,l)$ there is a region for which a
non-constant, soliton-like transition state (the ``instanton''
state) exists and has lower energy than the constant transition state. An
important assumption made in~\cite{martens1} is that the magnetostatic
contribution is dominated by the surface magnetostatic effect. This is
supported by the above simulations.  One would expect that by increasing
the ring width the bulk magnetostatic energy would increase in magnitude
and eventually become nonnegligible, so that other transitions paths become
more probable (e.g., vortex-like states). This should eventually occur;
however, up to the appreciable $\Delta R/R$ values studied here, the
approximations made in~\cite{martens1} remain valid.
\vskip 12pt 

This research is supported in part by U.S.~National Science
Foundation Grants DMR-0706322~(ADK), DMS-0601179 and 0651077~(DLS), and an
NYU~Research Challenge Fund award.


\vskip 10 pt
\begin{references}
\bibitem{martens1} K. Martens, D.L. Stein and A.D. Kent, Phys. Rev. B {\bf 73}, 054413 (2006).
\bibitem{donahue1999} M. J. Donahue and D. G. Porter, OOMMF User's Guide, Version 1.0, Interagency Report NISTIR 6376, National Institute of Standards and Technology, Gaithersburg, MD (Sept 1999). http://math.nist.gov/oommf/ 
\bibitem{yang2007} T. Yang, M. Hara, A. Hirohata, T. Kimura and Y. Otani, Appl. Phys. Lett. {\bf 90}, 022504 (2007).
\bibitem{castano2003} F.J. Casta\~no, C. A. Ross, C. Frandsen, A. Eilez, D. Gil, H. Smith, M. Redjal, and F.B. Humphrey. Phys. Rev. B {\bf 67}, 184425 (2003).
\bibitem{castano2006} F.J. Casta\~no, D. Morecroft, and C. A. Ross, Phys Rev B {\bf 74}, 224401 (2006).
\bibitem{moneck2006} Matthew T. Moneck and Jian-Gang Zhu, J. Appl. Phys. {\bf 99} 08H709 (2006).
\bibitem{rothman2001} J. Rothman, M. Kl\"aui, L. Lopez-Diaz, C.A.F. Vaz, A. Bleloch, J.A.C. Bland, Z. Cui and R. Speaks, Phys. Rev. Lett. {\bf 86}, 1098 (2001).
\bibitem{zhu2000} Jian-Gang Zhu, Youfeng Zheng and Gary A. Prinz, J. Appl. Phys. {\bf 87}, 6668 (2000).

\end{references}
\end{document}